\documentclass[amsmath,amssymb,preprint,nofootinbib,floatfix,superscriptaddress]{revtex4}
\usepackage[final,dvips]{graphicx}
\usepackage{hyperref}
\usepackage{multirow}
\usepackage{dcolumn}
\usepackage{slashed}

\begin{document}

\preprint{Brown-HET-1586}

\title{Dynamics of the chiral phase transition from AdS/CFT duality}

\author{Gerald Guralnik}
\email{gerry@het.brown.edu}
\affiliation{Department of Physics, Brown University, Providence, RI 02906 USA}
\author{Zachary Guralnik}
\email{zach@het.brown.edu}
\affiliation{Department of Physics, Brown University, Providence, RI 02906 USA}
\author{Cengiz Pehlevan}
\email{cengiz@het.brown.edu}
\affiliation{Edmond and Lily Center for Brain Sciences, The Interdisciplinary Center For Neural Computation, The Hebrew University of Jerusalem, Jerusalem, 91904 Israel}
\affiliation{Department of Physics, Brown University, Providence, RI 02906 USA}

\today

\begin{abstract}
We use Lorentzian signature AdS/CFT duality to study a first order phase transition in
strongly coupled gauge theories which is akin to the chiral phase transition in QCD. We discuss the relation between the
latent heat and the energy (suitably defined) of the component of a D-brane
which lies behind the horizon at the critical temperature.  A numerical simulation of a dynamical phase transition in an expanding, cooling Quark-Gluon plasma
produced in a relativistic collision is carried out.
\end{abstract}

\maketitle

%\tableofcontents

\section{Introduction}

At present, dual string theory descriptions of a number of large N gauge theories are known and facilitate computations at large 't Hooft coupling via AdS/CFT duality \cite{Maldacena97,Witten98,Gubser98}. Some of these are distant\footnote{Or close, depending on which question is being asked} cousins of QCD, and capture some of the qualitative features of strong coupling phenomena in QCD. Particular attention has been given to the phase structure of these theories \cite{Filev07,Filev071,Albash07,Erdmenger07,Filev09,Filev091,Evans101,Evans102} and the behavior under conditions arising in heavy ion collisions \cite{Janik10,Solana11}.  A prototype of the chiral phase transition in QCD is a first order phase transition in a strongly coupled large N ${\cal N}=2$ supersymmetric gauge theory, discovered in \cite{Babington03,Kruczenski04,Kirsch04,Apreda05,Albash06,Mateos06,Parnachev06,Aharony06} using the dual AdS description. In the Euclidean description, this transition is realized as a change in topology of space filling D7-branes in an asymptotically AdS--black hole background, and corresponds to a jump in a $\bar\psi\psi$ condensate.  A background dual to a theory more like QCD is described in \cite{Sakai04}, in which case the analogous change in topology corresponds to a chiral symmetry breaking transition.

In this article, we revisit the phase transition in the AdS dual of the prototype ${\cal N}=2$ supersymmetric model from a Lorentzian signature perspective, with an eye toward studying time dependent non-equilibrium processes. Non-equilibrium processes, such as the expansion and cooling of a quark-gluon fireball produced in a heavy ion collision, are not amenable to lattice gauge theory methods, but may be approached via a Lorentzian signature AdS/CFT duality.
Furthermore, there are very interesting questions about the interpretation of thermodynamic quantities in the dual gravitational description which are better asked in Lorentzian signature.  While quantities such as entropy and latent heat can be computed semi-classically from the Euclidean gravity/D-brane action, they lack a satisfactory physical interpretation\footnote{The Euclidean results correspond to an assumed saddle point approximation to integrals over degrees of freedom (e.g. string fields) which are not specified or even completely understood}.

In Euclidean signature, the boundary of the AdS-Schwarzchild black hole has topology $S^5\otimes S^1\otimes R^3$ which, in the ${\cal N}=2$ model, is wrapped by D7-branes having the boundary topology $S^3 \otimes S^1\otimes R^3$. The $S^1$ of the space-time contracts to zero size in the interior, at a point corresponding to the horizon after continuation to Lorentzian signature. In the low temperature phase the D7-brane ends smoothly in the interior by contraction of the $S^3$ wrapping the space-time $S^5$.  In the high temperature phase the D7-brane ends smoothly at the same point where the $S^1$ of the Euclidean space-time, which the D7-brane wraps, contracts to zero size.

Upon continuation to Lorentzian signature, the D7-brane in the low temperature phase ends before reaching the horizon, while the D7-brane extends through the horizon in the high temperature phase.  We will solve the static equations of motion for the D7-brane in the in-falling case in Section \ref{static}, including the region behind the horizon, where it can be shown that the D7-brane ends at a conical singularity.  Since the AdS radius becomes time-like inside the horizon, the conical singularity can be viewed as one-half of an annihilation diagram: the D7-brane ends before reaching the black-hole singularity by annihilation into closed-string modes. We show that the conical singularity corresponds to a spatial $S^1$ collapsing at the speed of light.  In passing dynamically from the high temperature phase to the low temperature phase,  the D7-brane can be said to ``pull out'' of the black hole, although what in fact is happening is that the component of the D7-brane  interior to the horizon vanishes by annihilation into gravitons, etc.

While for many questions the component of the D7-brane interior to the horizon  is irrelevant, it is presumably related to the existence of latent heat in the AdS-description of the phase transition.  In the low temperature phase the free energy, obtained from the Euclidean action, can be shown to be equivalent to a suitably defined energy of the D7-brane in the Lorentzian signature description, while in the high temperature phase the free energy corresponds to the energy of the component of the D7-brane lying outside the horizon.  The free energy is a bound on the energy available for work, which in this case corresponds to flux of energy across the AdS boundary.  This flux is bounded by the energy of the component of the D7-brane lying outside the horizon.  The flux of energy across the AdS boundary is defined only after a suitable renormalization, discussed in Section \ref{energy}. In a static AdS-black hole background, and neglecting back-reaction, there is conserved energy-momentum tensor associated with time translation invariance\footnote{Inside the horizon, the variable in which the metric is translation invariant becomes space-like.}, such that passing between two configurations with the same free energy requires flux across the AdS boundary match flux across the horizon.  The energy of the component of the D7-brane inside the horizon in the high temperature phase would therefore seem to be related to the  amount of energy required for the transition from low temperature (non-infalling) to the high temperature (infalling) phase, i.e. the latent heat.  We evaluate the energy of the component of the D7-brane interior to the horizon at the critical temperature in section \ref{latent}, finding a result which is about $1/8$ of the latent heat determined by the discontinuity in the $T\partial_T{\cal F}$ across the phase transition.  We do not yet have a satisfactory understanding of the discrepancy between the interior energy and the latent heat, although it is presumably related to the neglected back reaction.

With an eye towards simulating a strongly coupled expanding-cooling plasma, we construct a metric dual to an expanding cooling plasma in section \ref{dynamic}.
Similar metrics have been constructed previously; the initial attempts \cite{Janik05}  having a singularity where there should have been a horizon.  This problem was rectified in \cite{Heller08,Kinoshita08}, using Eddington-Finkelstein type coordinates.  We find it much more convenient to use a Painleve-Gullstrand or `river model' type metric.
%%which has the added advantage of asymptotically resembling a standard AdS metric,
%%related to the Eddington-Finkelstein coordinates by a coordinate transformation.
In section \ref{dynamicsect}, we describe our initial efforts to simulate the dynamics of a D7-brane passing through the first order phase transition in this background.

The work presented here appeared before in one of the authors' PhD thesis \cite{Pehlevan10}. A related paper \cite{Evans10} appeared as we were finalizing this work.

\section{Equilibrium thermodynamics in Euclidean signature}

%Thermodynamics of the D7-brane is calculated in Euclidean signature.  In this picture, the space ends at the black hole horizon and the Euclidean time direction has a periodicity that is proportional to the Hawking temperature of the black hole. For out-of-equilibrium processes, one needs to go to Lorentzian signature.

We consider the ${\cal N}=2$ super Yang-Mills theory obtained
by coupling an ${\cal N}=4$ multiplet to $N_f$
hypermultiplets in the fundamental representation of the gauge group
$SU(N_c)$. This theory is conformal in the limit  $N_c\rightarrow\infty$ with $N_f$
fixed\footnote{For finite $N_c$, this
theory is not asymptotically free and requires an ultraviolet
completion.}.  For large t'Hooft coupling $\lambda = g^2N_c >>1$, the theory is dual to weakly coupled  supergravity in an Anti-deSitter background with space-filling D7-branes \cite{Karch02}. The phase diagram and spectroscopy of this theory has
been studied in great detail using AdS/CFT duality. Of particular
note is a first order phase transition as the ratio of temperature to quark mass is varied.  This transition is akin to
the chiral phase transition in QCD\cite{Babington03,Kirsch04,Albash06,Hoyos07}.

Equilibrium thermodynamic quantities in this theory can be computed
from the dual string theory description of this theory, which
involves a background with a Euclidean black hole in $\text{AdS}^5\times
S^5$ space time,
\begin{align}\label{EucBH}
ds^2 = \frac{1}{z^2}\left( (1-b^4z^4)dt^2 + \frac{1}{1-b^4z^4}dz^2 +
d\vec x^2\right) +  d\Omega_5^2\, ,\end{align} where it is convenient to
write the five-sphere metric as
\begin{align}\label{five}
d\Omega_5^2 = d\theta^2+\cos^2\theta(d\phi^2 + \cos^2\phi d\Omega_3^2)\, .
\end{align}
Throughout this paper, we will set the AdS curvature, $R$, to $1$. This space-time is defined for $z \le 1/b$ and is smooth and
complete if the Euclidean time $t$ is compactified on a circle of
radius $\pi/b$,  corresponding to the inverse temperature.

The inclusion of the $N_f$ hypermultiplets (quarks) corresponds to
the addition of  $N_f$ D7-branes to this background, embedded on a
surface
\begin{align}
\phi = 0, \qquad \theta = \theta(z).
\end{align}
The induced metric on the D7-brane is
\begin{align}
ds^2 = \frac{1}{z^2}\left( (1-b^4z^4)dt^2 + \frac{1}{1-b^4z^4}dz^2 +
d\vec x^2\right) + \theta'(z)^2 dz^2 + \cos^2\theta dS_3^2\,
,\end{align}
giving the D7-brane action
\begin{align}\label{EucAction}
I_{\text{D7}}=-{\cal N}\beta\int_{z_0^{\rm horizon}} dz \frac{\cos^3\theta}{z^5}\sqrt{1+z^2(1-b^4z^4)\theta'^2}.
\end{align}
with ${\cal N}=N_f T_{\text{D7}}\Omega_3\left(\int d^3x\right)$, where $T_{\text{D7}}$ is the brane tension and $\Omega_3$ is the volume of the three-sphere. The action \eqref{EucAction} is divergent as the boundary of integration $z\to0$ and needs to be renormalized, which will be discussed in more detail below.
This gives rise to the equations of motion
\begin{align}\label{staticEOM}
0=&z^2(-1+b^4z^4)\theta'' + z(3+b^4z^4)\theta'+
2z^3(1-b^4z^4)(2-b^4z^4)\theta'^3 \nonumber \\
&+3\tan\theta\left(-1+z^2(-1+b^4z^4)\theta'^2\right) \, .
\end{align}

Near the AdS boundary, $z\rightarrow 0$, the solutions have the
asymptotic behavior
\begin{align} \theta(z) \sim\theta_1 b z + \theta_3 b^3z^3+ \cdots\,
,\end{align}
where the parameters $\theta_{1,3}$ determine the quark
mass $M$ and chiral condensate $C$.  Specifically \cite{Mateos07},
\begin{align}\label{CandM}
&M = \frac{1}{2}\sqrt{\lambda} T \theta_1,\\
&C \equiv \left<q\bar q\right> = -\frac{1}{8}N_fN_c T^3 \sqrt{\lambda}\left(-2\theta_3
+ \frac{\theta_1^3}{3}\right).\end{align}

\begin{figure}
\includegraphics[width=0.5\textwidth]{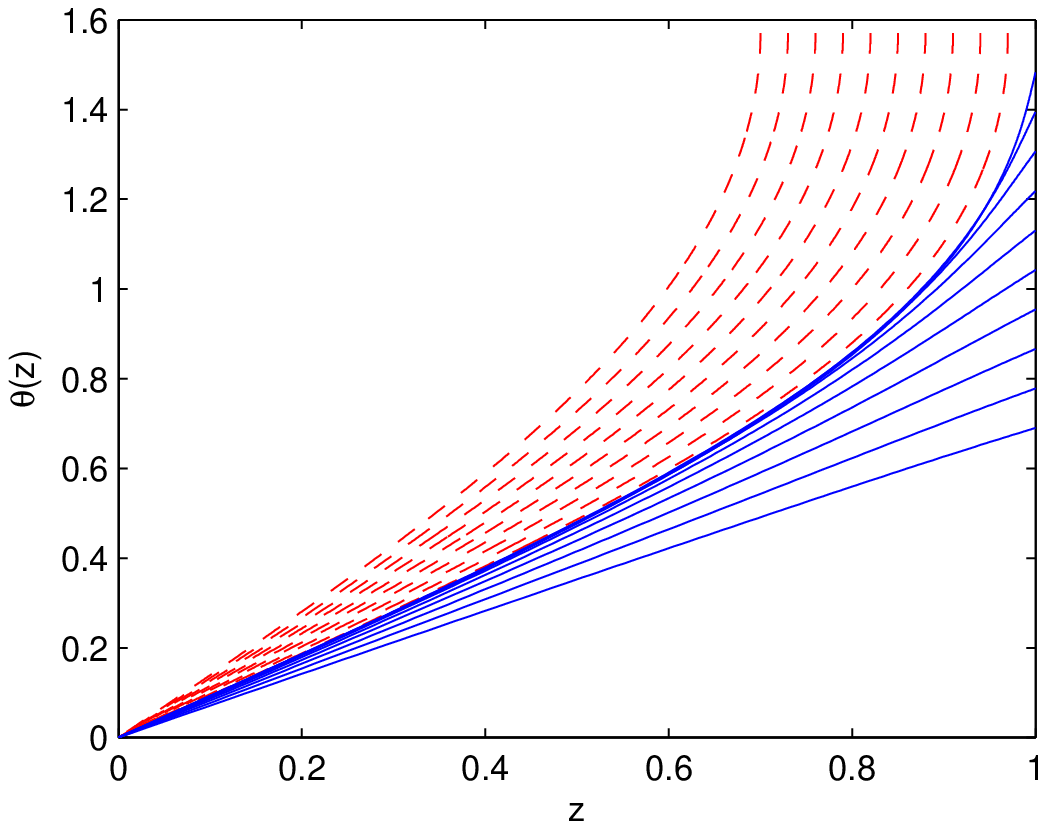}\\
\includegraphics[width=0.45\textwidth]{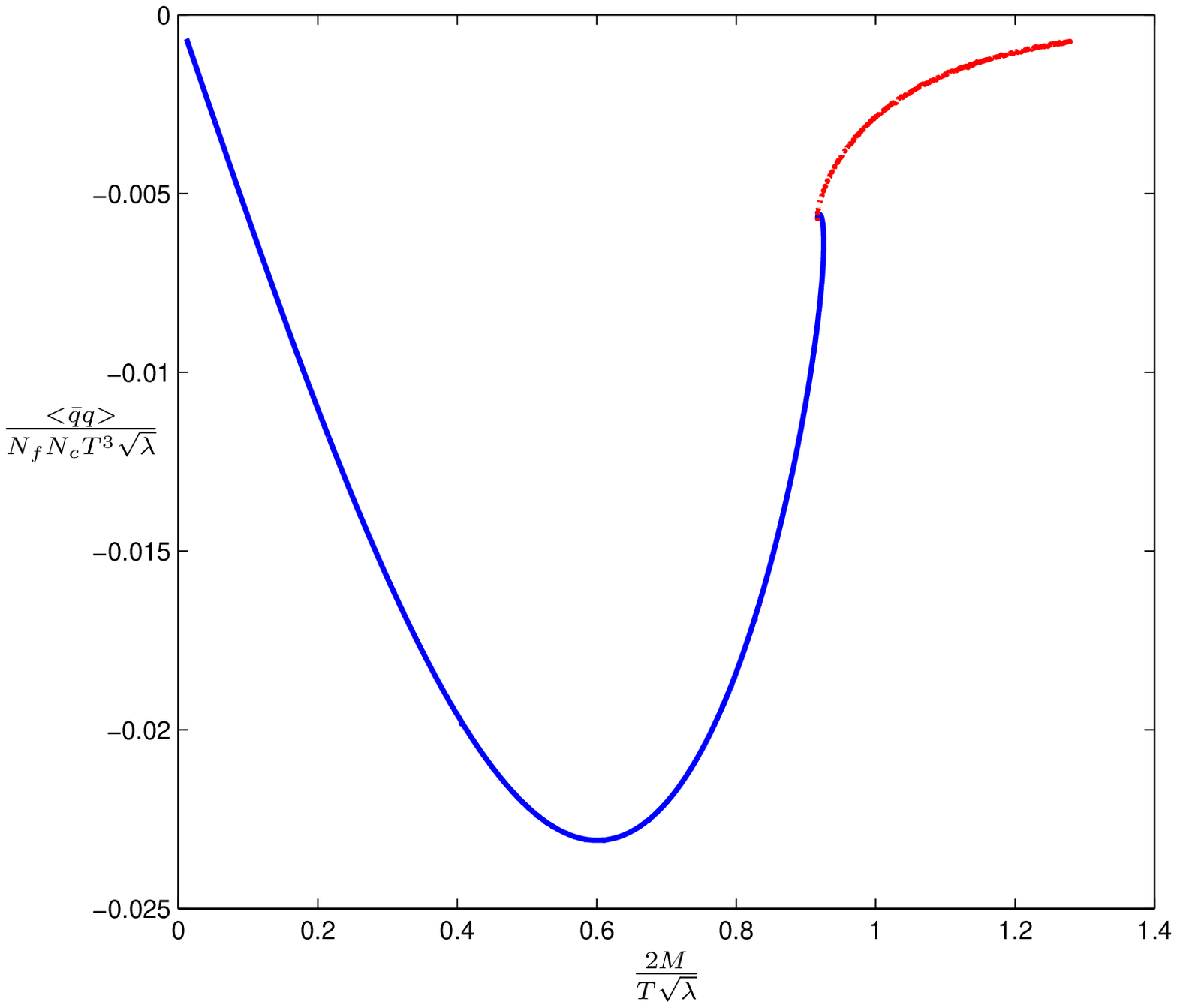}
\includegraphics[width=0.45\textwidth]{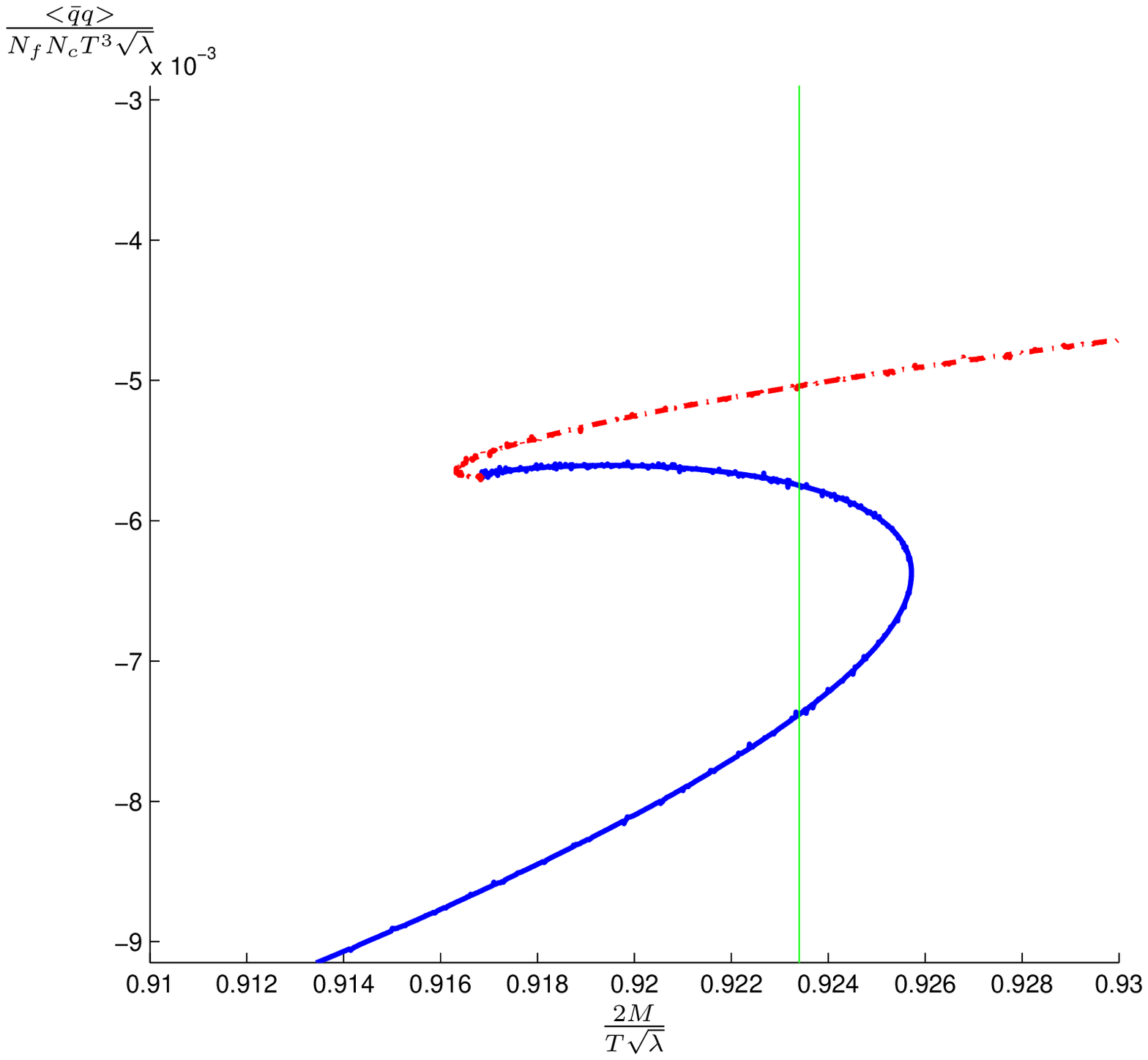}
\caption{\label{staticEuclidean}On top, D7-brane embeddings in the Euclidean black hole background is shown. $b$ is set to 1. Blue (solid) lines plot embeddings that extend to the horizon and red (dashed) lines plot embeddings that do not extend to the horizon. Below, quark condensate as a function of M/T that arises from the D7-brane embeddings is shown. The figure on bottom right zooms into the multivalued region of the figure on bottom left. The green (vertical) line shows the critical mass at
which the condensate’s vev jumps discontinously, at which $\frac{2M}{T\sqrt\lambda}=0.9234$. This value
was found by equating the area between the green (vertical) line and the C–M curve
on both sides of the green (vertical) line \cite{Albash06}.}
\end{figure}

The topology of a constant $z$ slice of the D7-brane at the AdS boundary
$z\rightarrow 0$ is $R^3\otimes S^1 \otimes S^3$. The D7-brane can end
at $z_{\rm end} <1/b$ if the $S^3$ contracts to zero size, with
$\theta({\rm z_{\rm end}})=\pi/2$, and $\theta'(z_{\rm end})
=\infty$ so that the end is smooth rather than a conical
singularity.  If on the other hand $z_{\rm end} =1/b$,  the $S^1$ of
the space-time in which the D7-brane is embedded contracts to zero
and the boundary condition $\theta'(z_{\rm end}) = \frac{3}{4}\tan\theta(z_{\rm end})$ follows
from the equations of motion \eqref{staticEOM}. Solving the D7-brane
equations of motion at fixed temperature, subject to the boundary
conditions at $z_{\rm end}$, yields a curve in the $C$-$M$ plane
parameterized by $z_{\rm end}$ for D7-branes which do not extend to
the horizon and $\theta(z_{\rm horizon})$ for D7-branes which end at
the horizon (see figure \ref{staticEuclidean}). For these embeddings, the free energy $F$ is given by $\beta I_{\text{D7}}$,
and the free energy density ${\cal F}=F/\int d^3x$ satisfies
\begin{align}\label{df}
d{\cal F}=%-{\cal S}dT=CdM,
CdM
\end{align}
at constant temperature.
In a certain range of values for $M$
there are solutions with several values of $C$, indicating a first order phase transition. The transition occurs at a critical value of the only dimensionless parameter $T/M$.
At a given temperature, the critical mass can be determined by Maxwell's equal area construction \cite{Albash06}. It lies the a point $M=M_c$ where there are three solutions, such the integral over the curve $C(z_{\rm end}), M(z_{\rm end})$ connecting them gives $\int C dM =0$. The physical solutions at the critical point are the outer two which for which the free energy is convex, $\frac{d^2f}{dM^2}=\frac{dC}{dM}>0$, and takes the same value via the equal area rule.  Physical transitions from one phase to the other do not follow the curve $C(z_{\rm end}), M(z_{\rm end})$, but occur by bubble nucleation or supercooling/heating. The solutions we describe subsequently focus on supercooling/heating which is more numerically tractable.

%At zero temperature,  the theory has a moduli space of vacua,
%including Coulomb, Higgs and mixed Coulomb-Higgs vacua, whose AdS
%description has been given in \cite{?}. It has been shown in
%\cite{??} that the moduli space is largely lifted at finite
%temperature.  The Higgs and mixed Coulomb-Higgs vacua are
%characterized by instantons on the D7-branes with gauge connections
%in the $S^3$ and $z$ directions. Numerical calculations in \cite{?}
%indicate that at temperatures below the critical temperature,
%instantons are driven toward zero size, the point at which the Higgs
%or mixed Coulomb-Higgs branch meets the Coulomb branch, while
%instantons are driven towards finite size above the critical
%temperature.  In fact, the mechanism by which this occurs is closely
%related to the mechanism responsible for chiral symmetry restoration
%in the AdS dual of a more QCD-like theory (This point will be
%elaborated upon later).

\section{Static D7-brane embedding in Lorentzian signature\label{static}}

Before considering  D7-branes embedded in the time dependent
background \eqref{our}, we will discuss some of the properties of
the static equilibrium solution in Lorentzian signature.
%Most of the
%discussion of the equilibrium case in the literature is in Euclidean
%signature,  with the exception of some spectral
%computations\cite{??}, and there has been no analysis of the
%behavior of the D7-branes behind the horizon.
While the static embedding outside the horizon is the
same in either Euclidean or Lorentzian signature, it is interesting
to consider the behavior of the D7-branes behind the horizon, a region which exists only in Lorentzian signature. For reasons to be discussed later, the behavior of the D7-brane behind the horizon is related to the latent heat.

\begin{figure}
\includegraphics[width=\textwidth]{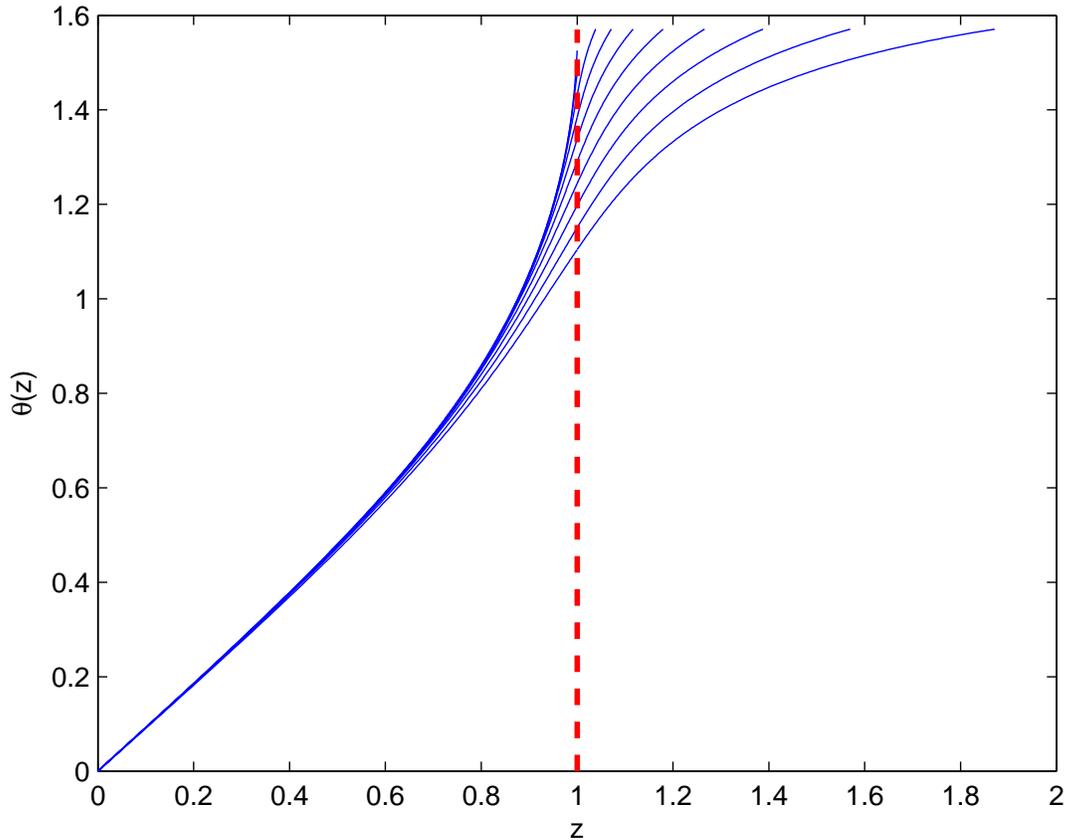}
\caption{\label{staticLorentzian} Static D7-brane embedding in Lorentzian signature. $b$ is set to 1.}
\end{figure}

The Lorentzian signature AdS-Schwarzchild black hole metric is
\begin{align}
\label{LorBH}
ds^2 = \frac{1}{z^2}\left( -(1-b^4z^4)dt^2 + \frac{1}{1-b^4z^4}dz^2 +
d\vec x^2\right) +  d\Omega_5^2\, .
\end{align}
A redefinition of the time coordinate,
\begin{align}
t_r=t-\int^z \frac{b^2z'^2}{1-b^4z'^4}dz'\, ,
\end{align}
gives a metric of the Painlev\'e--Gullstrand, or 'river-model' form
\begin{align}\label{PainGull}
ds^2 = \frac{1}{z^2}\left(-dt_r^2 + (dz - b^2 z^2 dt_r)^2+ d\vec
x^2\right) + d\theta^2+ \cos^2\theta(d\phi^2 + \cos^2\phi
dS_3^2)\, ,\end{align}
which has the advantage that there is no coordinate singularity at the horizon. Moreover, time dependence can later be introduced in the parameter $b$ without generating a curvature singularity. Since only the time coordinate has been redefined, the equations for $\theta(z)$ for a static D7-brane embedding are
identical to those in equation \eqref{staticEOM}. The only difference from
the solutions in the Euclidean background \eqref{EucBH} is that the
solutions which reach the horizon may be continued behind the
horizon to $z>1/b$. This is done by enforcing continuity across the horizon, see Figure \ref{staticLorentzian}. Numerically,  one finds that the infalling
solutions reach $\theta = \pi/2$ at a finite value of $z
>1$.  Thus, the infalling D7-branes end like the non-infalling D7-branes, via an $S^3$
contracting to zero size.  However the ending is not smooth due to
causality.  Inside the horizon $z$ is timelike, so the boundary
condition required for smoothness, $\theta'(z_{\rm end})=\infty$,
would correspond to an $S^3$ collapsing faster than the speed of
light. Upon inspection of \eqref{staticEOM}, the correct boundary
condition for infalling solutions is seen to be
\begin{align}
-1+z_{\rm end}^2(-1+b^4z_{\rm end}^4)\theta'(z_{\rm end})^2 =0\, ,
\end{align}
which corresponds to an $S^3$ collapsing at the speed of light. A suitably defined energy associated with the cone lying behind the horizon would seem to be a natural guess for the latent heat\footnote{'Latent' means hidden, which in this case has a very literal realization as hidden behind the horizon}. We will explore this proposal more deeply in the subsequent discussion.

\section{Remarks on Free Energy \label{energy} and the Stress-Energy Tensor}

In this section we calculate the renormalized stress-energy tensor of the D7-brane. We will see that the free energy is equivalent to energy of the component of the static D7-brane lying outside the horizon. The free energy represents a bound on mechanical work which can be obtained from a thermodynamic system at constant temperature. In the present case, the energy external to the horizon represents a bound on energy flux at the AdS boundary. We will also compute the component of the D7-brane stress tensor determining the energy flux at the AdS-boundary boundary and horizon-crossing.  For quasi-static time evolution, we will see that the flux at the AdS boundary is
$C\dot M$, consistent with \eqref{df}.  Although $\int C dM$ vanishes along the curve of (not always stable) solution connecting the two phases which coexist at the critical temperature, the physical passage between phases does not occur this way, nor is it quasi-static, such that the net flux at infinity may be non-zero. In a transition between these phases which satisfies the equations of motion, there is a lower bound on the D7 energy flux at infinity determined by the energy of the cone interior to the horizon in the infalling phase.  A lower bound on flux at the AdS boundary seems clearly related to the existence of latent heat. However we will see that the cone energy accounts for only a fraction of the latent heat.

\subsection{Renormalized action}

The Euclidean action of the D7-brane, given by \eqref{EucAction} is divergent
 as $z\to 0$, but rendered finite by holographic renormalization
 \cite{Skenderis02,Balasubramanian99,Karch05} .
 Necessary counterterms to renormalize the action were given in \cite{Karch05}. One introduces a finite cut-off at $z=\epsilon$ slice and adds the counterterms made of curvature invariants at the $z=\epsilon$ slice. The authors of \cite{Karch05} calculated these counterterms in Fefferman-Graham coordinates. The counterterms are given by
\begin{align}\label{counterterms}
	L_1 &=- \frac{1}{4}\sqrt{\gamma}, \nonumber \\
	L_2 &= -\frac 1{48} \sqrt{\gamma}\, R_\gamma, \nonumber\\
	L_3 &= -\ln u_\epsilon\, \sqrt{\gamma}\,\frac 1{32}\left(R_{ij}R^{ij}-\frac 13 R^2_\gamma \right), \nonumber \\
	L_4 &= \frac 12 \sqrt{\gamma}\,\theta(x,\epsilon)^2, \nonumber \\
	L_5 &=-\frac 12 \ln \epsilon \, \sqrt{\gamma}\,\theta(x,\epsilon)\left(\Box_\gamma+\frac 16 R_\gamma \right)\theta(x,\epsilon) \nonumber\\
	L_F &= -\frac 5{12}\sqrt{\gamma}\,\theta(x,\epsilon)^4.
\end{align}
$\gamma_{ij}$ is the induced brane metric on the cut-off slice (at coordinate $\epsilon$ in the radial direction) in these coordinates.
We denote the renormalized action by $I_{\rm ren}$.
In \cite{Grosse07}, the D7 brane action renormalization was done for a time-dependent embedding. The free energy is given by ${F}=I_{\rm ren}/\beta$. In Lorentzian signature, this turns out to be equivalent to energy (suitably defined) of the component of the D7-brane outside the horizon.

%For the static AdS black hole background,
%
%\begin{align}
%ds^2 = \frac{1}{z^2}\left(- (1-b^4z^4)dt^2 + \frac{1}{1-b^4z^4}dz^2 +
%d\vec x^2\right) +  dS_5^2
%\end{align}
%
%there is an isometry, translation of $t$, associated with the Killing vector $\xi^A\partial_A=\frac{\partial}{\partial t}$ from which one can define a conserved energy-momentum tensor of the D7-brane,
%\begin{align}
%\partial_A(\xi^B\sqrt{-g}T^A_B)=0
%\end{align}
%
%where
%
%\begin{align}
%T^{AB}(x) &= \frac{2}{\sqrt{-g}}\frac{\delta I_{D7}}{\delta g_{AB}}.
%\end{align}
%
%Evaluating the energy outside the horizon
%
%\begin{align}
%E=\int_{z_{horizon}}^0 dz dx^3 d\cdots T^t_t
%\end{align}
%
%yields a divergent result which is again canceled by counterterms. The counterterms can be determined by For the static D7-brane solutions, evaluating the energy outside the horizon, yields a result equivalent the free energy computed from the Euclidean signature action,
%\begin{align}
%{\rm result}
%\end{align}
%
%Since free energy is a bound on energy available for work,  and corresponds to the energy outside the horizon, it is natural to identify work as a flux of energy across the AdS boundary $r=\infty$. For a non-static D7-brane,  the flux across the boundary is
%\begin{align}
%\int_{z=\epsilon} T^t_z\cdots...
%\end{align}
%which is again divergent and renormalized as follows:
%--DISCUSSION OF TIME DEPENDENT COUNTERTERMS--
%yielding
%\begin{align}
%\int_{z=\epsilon} T^t_z\cdots... = m\dot c
%\end{align}

\subsection{Stress energy tensor}

The Hilbert stress energy tensor is
\begin{align}
	T^{AB}(x) &= \frac{2}{\sqrt{-g}}\frac{\delta I_{D7}}{\delta g_{AB}}.
\end{align}
Given a killing vector $\xi^{A}$, one can define conserved currents
\begin{align}\label{conservation}
	\partial_A \left(\xi^B\sqrt{- g}\,T^A_{\,\,\,B}\right)=0.
\end{align}
We consider a time dependent embedding, $\theta=\theta(z,t)$, on the Lorentzian AdS-black hole background,
\begin{align}\
ds^2 = \frac{1}{z^2}\left(- (1-b^4z^4)dt^2 + \frac{1}{1-b^4z^4}dz^2 +
d\vec x^2\right) +  d\Omega_5^2\, ,
\end{align}
 and the Killing vector $\xi^A\partial_A=\frac{\partial}{\partial t}$.  We will call these coordinates "Schwarzschild coordinates" in the following. Then, in
 the absence of counterterms,
\begin{align}
  &-\frac{\sqrt{-g}\,T^t_{\,\,\,\,t}}{N_fT_{D7} } = \delta(\psi)\delta(\theta - \Theta(z,t)) \det{\Omega_3}\frac{\cos^3\theta}{z^5}\frac{1+z^2(1-b^4z^4)\theta'^2}{\sqrt{1+z^2(1-b^4z^4)\theta'^2-\frac{z^2\dot\theta^2}{1-b^4z^4}}} \nonumber \\
  &-\frac{\sqrt{-g}\,T^z_{\,\,\,\,t}}{N_fT_{D7}} =\delta(\psi)\delta(\theta - \Theta(z,t)) \det{\Omega_3}\frac{\cos^3\theta}{z^5}\frac{-z^2(1-b^4z^4)\theta'\dot\theta}{\sqrt{1+z^2(1-b^4z^4)\theta'^2-\frac{z^2\dot\theta^2}{1-b^4z^4}}}.
\end{align}
Here $\det{\Omega_3}$ refers to the determinant of the three-sphere part of the metric \eqref{five} and $\Theta(z,t)$ is the classical trajectory. As expected, for static configurations $\sqrt{-g}\,T^z_{\,\,\,\,t}$ vanishes. One can then define a conserved energy associated with the probe D7-brane,
\begin{align}\label{UD7}
U_{D7}&= -\int dz d^3x d\ldots\, \sqrt{-g}\,T^t_{\,\,\,\,t} ={\cal N}\int dz \, \frac{\cos^3 \Theta}{z^5}\frac{1+z^2(1-b^4z^4)\Theta'^2}{\sqrt{1+z^2(1-b^4z^4)\Theta'^2-\frac{z^2\dot\Theta^2}{1-b^4z^4}}}.
\end{align}
Thus, when the embedding is static, the conserved energy of the part of the D7 brane outside the horizon
is formally equal to the brane's free energy (see \eqref{EucAction}).
%This is suspicious, because the thermodynamic internal energy (density) ${\cal U}={\cal F} + T{\cal S}$, and is certainly not equal to ${\cal F}$. The answer lies in realizing that, as pointed out in \cite{Karch09}, the thermodynamic internal energy has contributions due to the backreaction of the brane. The energy satisfies
%
%\begin{align}
%&\frac{dU_{D7}}{dt} = \int d^3x d\ldots\, \left.\sqrt{-g}\,T^z_{\,\,\,\,t}\right|_{z=z_{\text{end}}} - \int d^3x d\ldots\, \left.\sqrt{-g}\,T^z_{\,\,\,\,t}\right|_{z=z_{\text{0}}}.
%\end{align}
%
%Recalling that $-\sqrt{-g}\,T^z_{\,\,\,\,t}$ is the energy flow through a constant $z$ surface in the positive $z$ direction, the equation above is poining that the increase in energy is given by the flow of energy from both ends of the brane. Later, we will be interested in the energy that flows into the horizon at the tip of the brane. This will require us to look at $-\sqrt{-g}\,T^z_{\,\,\,\,t}$ at $z_{\text{end}}=1/b$.

The stress-energy tensor as defined above is divergent as $z\to 0$ and leads to divergent energies.
Including the counterterms \eqref{counterterms} in the renormalized action
yields the canonical stress energy tensor components\footnote{The Hilbert stress tensor computed with the renormalized action does not yield finite
energy unless the configuration is static.}
%WHY!!!!}
%
\begin{align}\label{renormalizedHam}
&-\frac{{\cal T}^t_t}{N_fT_{D7} } =\delta(\psi)\delta(\theta - \Theta(z,t)) \det{\Omega_3}\frac{\cos^3\theta}{z^5}\frac{1+z^2(1-b^4z^4)\theta'^2}{\sqrt{1+z^2(1-b^4z^4)\theta'^2-\frac{z^2\dot\theta^2}{1-b^4z^4}}} \nonumber\\ &\qquad
-\lim_{\epsilon\to 0}\left\lbrace
\delta(z-\epsilon)\delta(\psi)\delta(\theta-\Theta(t,z))\delta^3(\ldots)\Omega_3 \sqrt{-\gamma} \left[\frac 14-\frac 12\theta^2+\frac 5{12}\theta^4+ \frac{1}{2}\ln u_{\epsilon}\,\gamma^{tt}(\partial_t\theta)^2\right]\right\rbrace.
\end{align}
where $\gamma_{ij}$ is the induced metric on $z=\epsilon$ slice.
Integrating \eqref{renormalizedHam} over $z$ yields a finite Hamiltonian, $H=\int_\epsilon\, dz {\cal T}^t_t$.
There is a conservation law
\begin{align}
\partial_t{\cal T}^t_z - \partial_z {\cal T}^z_t,
\end{align}
which can be used to renormalize ${\cal T}^z_t$ as $z\to0$
\begin{align}
 &\lim_{z\to0}-\frac{{\cal T}^z_{t}}{N_fT_{D7}} =\lim_{z\to0}\left\lbrace \delta(\psi)\delta(\theta - \Theta(z,t)) \det{\Omega_3}\frac{\cos^3\theta}{z^5}\frac{-z^2(1-b^4z^4)\theta'\dot\theta}{\sqrt{1+z^2(1-b^4z^4)\theta'^2-\frac{z^2\dot\theta^2}{1-b^4z^4}}}\right.\nonumber \\
&\qquad\left. -\delta(\psi)\delta(\theta-\Theta(t,\epsilon))\delta^3(\ldots)R^4\Omega_3 \sqrt{-\gamma} \left[-\theta\dot\theta+\frac 5{3}\theta^3\dot\theta+ \ln u_{\epsilon}\,\gamma^{tt}\dot\theta\ddot\theta\right]\right\rbrace,
\end{align}
and to show that the flow of energy at the AdS boundary is
\begin{align}\label{flow}
&-\lim_{z\rightarrow 0}\frac{{\cal T}^z_t}{N_fT_{D7}} =\delta(\psi)\delta(\theta-\Theta(t,\epsilon))\delta^3(\ldots)\Omega_3\left[\frac {b^4}{3}\dot\theta_1\theta_1^3-2b^4\dot\theta_1\theta_3-\frac {b^2}{2}\dot\theta_1\ddot\theta_1\right].
\end{align}

In a quasi-static process, keeping only terms which are first order in time derivatives,
\eqref{flow} and \eqref{CandM} implies
\begin{align}
\dot H = \int C \dot M\, ,
\end{align}
Note that \eqref{flow} assumes that there can only be flow from the AdS boundary and not the endpoint of the D7-brane.  This is true if the D7-brane ends (smoothly) before reaching the horizon.  However there may be flow across the singular endpoint of a D7-brane ending inside the horizon.
Note that the energy flux across the horizon is ${\cal O}(\dot\theta^2)$, and so vanishes in quasi-static evolution.  This is consistent with $dF=CdM$ where $F$, the free energy, is the component of the energy outside the horizon $\int_{z_{\rm horizon}}^0 {\cal T}^t_t$.

\subsection{Latent Heat \label{latent}}

The two phases which coexist at the critical temperature have the same free energy,  although the D7-brane Hamiltonian differs by the energy of the cone behind the horizon in the infalling phase.  Passing dynamically from one phase to the other is not a quasistatic process; the energy of the cone
\begin{align}\label{bound}
E_{\rm cone}=\int_{z_{\rm end}}^{z_{\rm horizon}} dz\, {\cal T}^t_t\, ,
\end{align}
is a bound on the flux of D7-brane energy across the horizon in passing from the non-infalling phase to the infalling phase.  Given the equivalence of the free energy on either side of the transition, this is also a bound on the flux across the AdS boundary.   As a bound on the energy which must be added to the system in a phase change at the critical temperature, and on $T\Delta S$ corresponding to the change in black hole mass, this would seem to be equivalent to the latent heat.
However this is not the most stringent bound.
The latent heat can be reliably computed from the discontinuity in the derivative
of the free free energy with respect to temperature,
\begin{align}\label{lat}
Q_{\rm latent}&=T\Delta S=-T\Delta\left(\frac{\partial F}{\partial T}\right)\\
&= M\Delta C= 0.0011\,\lambda T^4N_fN_c.
\end{align}
This result is about eight times larger than \eqref{bound}, which is computed numerically for the infalling solution at the critical temperature.
The discrepancy between \eqref{bound} and \eqref{lat} is presumably related to the fact that back-reaction has not been taken into account.  Due to the equations of motion, the effect of the back-reaction on the free energy is of smaller order in the $1/N_c$ expansion than the effect of the back reaction on the internal energy $U=F+TS$.  The effect of back-reaction on the latter is of the same order, $N_fN_c$, as the D7-brane energy \cite{Karch09}.  The latent heat may be reliably calculated from \eqref{lat}, while \eqref{bound} captures only a partial contribution.

\section{Dynamics of an expanding plasma\label{dynamic}}

One of advantages of AdS/CFT duality over
lattice gauge theory is the possibility of simulating time dependent, non-equilibrium processes in a strongly gauge theory.  In the subsequent discussion we describe initial efforts at a numerical simulation of the dynamical passage through the phase transition in an expanding cooling plasma.  Much of this work has already appeared in \cite{Pehlevan10}.  As this article was being completed, a closely related article appeared \cite{Evans10}.

An approximate Lorentzian signature solution of Einstein's equations
dual to a boost invariant expanding cooling ${\cal N}=4$ Yang-Mills plasma,
akin to that arising in heavy ion collisions, was given in \cite{Janik05}.
In Fefferman Graham coordinates, the metric is
\begin{align}\label{JP}
ds^2 =
\frac{1}{\bar z^2}\left[-\frac{\left(1-\frac{e_0}{3}\frac{\bar z^4}{\bar \tau^{4/3}}\right)^2}{1+\frac{e_0}{3}\frac{\bar z^4}{\bar\tau^{4/3}}}d\bar\tau^2
+ \left(1 + \frac{e_0}{3}\frac{\bar z^4}{\bar\tau^{4/3}}\right)(\bar\tau^2d\eta^2
+ dx_\perp^2)\right] + \frac{d\bar z^2}{\bar z^2}\, ,
\end{align}
where $\bar\tau,\eta,x_\perp$ are the coordinates of the the Yang-Mills
theory and of the AdS boundary at $\bar z\rightarrow 0$.  These are the
natural coordinates for a relativistic heavy ion collision where,
for a collision along the $x^3$ axis, $\bar\tau^2 \equiv t^2 -(x^3)^2$,
$\eta =\frac{1}{2}\ln\frac{t-x^3}{t+x^3}$ (the rapidity), and
$x_\perp = x^1,x^2$.
%Homogeneity of the solution in $x_\perp$, as if
%the initial collision was between pancakes infinitely extended in
%$x_\perp$, is a reasonable assumption so long as one only considers
%the mid-rapidity region.
The asymptotic $\bar z\rightarrow 0$ expansion
of the metric \label{JPsoln} yields the expectation value of the
Yang-Mills theory stress energy tensor, via $g_{\mu\nu} =
g_{\mu\nu}^{(0)} + \bar z^4 <T_{\mu\nu}> + \cdots$.  The stress energy
tensor derived from \eqref{JP} corresponds to a relativistic
perfect fluid with boost invariant initial conditions \cite{Janik05},
with energy density
\begin{align} \label{BjH}
\varepsilon(\tau) = \frac{e_0}{\bar\tau^{4/3}} .
\end{align}
%

%The reasoning of \cite{Janik05} in deriving the metric \eqref{JP} was actually motivated by asymptotic behavior of D7-brane embeddings that we discussed above. They noticed that plugging a series expansion to the static equation of motion \eqref{staticEOM} did not fix $\theta_3$. Given a $\theta_1\sim M/T$, each $\theta_3$ defines a solution, however only one particular $\theta_3$ leads to a non-singular solution, which gives the physical value of the quark condensate. They applied this reasoning to the boundary metric and stress-energy tensor. They found that enforcing non-singularity one ended up with the time dependent metric \eqref{JP}, and the dual stress-energy tensor described Bjorken hydrodynamics \cite{Bjorken}, whose characteristic result is given by the $\bar\tau^{-4/3}$ fall in energy density, equation \eqref{BjH}.

The metric \eqref{JP} above is  the leading term in a late time expansion in terms of a scaling variable $s=\bar z\bar \tau^{-1/3}$, such that $g_{\mu\nu}=g_{\mu\nu}^{(0)}(s)+{\cal O}(\bar\tau^{-2/3})$. The dynamics of non-infalling D7-branes in this background was studied in \cite{Grosse07}, at temperatures well above the phase transition.
However \eqref{JP} is not well suited for studying the non-equilibrium dynamics of the phase transition, since it does not give a valid description of the region of space-time in the neighborhood of the horizon.
This problem was resolved in \cite{Heller08,Kinoshita08}, in which a solution was given in Eddington-Finkelstein coordinates.  However, we have found yet another form of the solution, involving a metric of the Painlev\'e-Gullstrand from, to be far more convenient.

\textbf{}
\subsection*{Proposed Background}

In this section we introduce our proposed background using phenomenological arguments. Later on we compare to the metric of Janik and Peschanski, \eqref{JP}, and show that both metrics are related by a coordinate transformation.
%We further do consistency checks on our metric, verify that it has an apparent horizon and calculate curvature invariants to check absence of naked singularities.
%
Before making our educated guess, let us note that the solution \eqref{JP} closely resembles an AdS-black hole in Fefferman-Graham coordinates,
\begin{align}\label{FG}
ds^2 = \frac{1}{\bar z^2}\left[-\frac{(1-b^4\bar z^4/4)^2}{1+b^4\bar z^4/4}dt^2 + (1
+ b^4\bar z^4/4)d\vec x^2\right] + \frac{d\bar z^2}{\bar z^2}\, ,
\end{align}
where the replacements $t\rightarrow \bar\tau$ and $d\vec x^2\rightarrow\bar\tau^2 d\eta^2 + d\vec
x_\perp^2$ are made to realize boost invariance in an expanding plasma, and $b$ is a function of $\bar\tau$,
\begin{align}\label{b}
b(\bar\tau)=\left(\frac{4e_0}{3}\right)^{1/4}\bar\tau^{-1/3}
\end{align}

We require a Lorentzian signature solution which is valid from the
boundary through the horizon.
Therefore, as a starting point consider the
following metric of a static AdS black hole,
\begin{align}\label{staticriver}
ds^2 = \frac{1}{z^2}\left(-dt_r^2 + (dz - b^2 z^2 dt_r)^2+ d\vec
x^2\right)
\end{align}
 where \eqref{staticriver} is related to the AdS-Schwarzchild metric \eqref{LorBH}
 by the coordinate transformation
\begin{align}\label{river}
t_r = t-\int^z\,\frac{b^2 z'^2}{1-b^4z'^4}dz'\, .
\end{align}
These coordinates are similar to the ``river model'' or
Gullstrand-Painleve coordinates of a Schwarzchild black hole in an
asymptotically Minkowski space \cite{Gullstrand22,Painleve21}.
To obtain a result consistent with an expanding plasma akin to one produced in a heavy ion collusion,
we replace the metric \eqref{river} with a boost
invariant version and allow $b$ to depend on $\tau$;
\begin{align}\label{our}
ds^2 = \frac{1}{z^2}\left(-d\tau^2 + (dz - b(\tau)^2 z^2 d\tau)^2+ \tau^2
d\eta^2 + d\vec x_\perp^2\right)\, .
\end{align}

The geometry \eqref{our} looks like a AdS black hole solution with horizon moving in the bulk according to $z_h=1/b(\tau)$. With an abuse of terminology, we call this surface a horizon, even though in the $\tau$ dependent case it is non-trivial to determine the existence and location of the horizon. For slowly varying
$b(\tau)$, the volume form on the horizon is
\begin{align}
{\cal A} = b(\tau)^3 \tau d\eta\wedge dx^2 \wedge dx^3.
\end{align}
The volume corresponds to an entropy per unit rapidity per unit transverse
area $S(\tau) = b(\tau)^3 \tau$, which must increase or remain
constant with increasing $\tau$.  If the expansion is isentropic,
then $b(\tau) \sim \tau^{-1/3}$.

With $b$ given by \eqref{b}, it is not hard to see that the following change
of coordinates
\begin{align}
&z = \bar z\frac{1}{\sqrt{1+\bar s^4/4}}, \qquad\tau =\bar\tau\left(1+\frac {\tau_2(\bar s)}{\bar\tau^{2/3}}\right),
\end{align}
where $\bar s=\left(\frac{4e_0}{3}\right)^{1/4}\bar z\bar \tau^{-1/3}$ and
\begin{align}
\tau_2(\bar s) = -\left(\frac{4e_0}{3}\right)^{-1/4}\int^s d\bar s'\, \frac{\bar s'^2}{\left(1+\bar s'^4/4\right)^{1/2}\left(1-\bar s'^4/4\right)}
\end{align}
transforms between our proposal and \eqref{JP}, to leading order in the $\bar\tau^{-2/3}$ expansion of \cite{Janik05}.

The advantage of the metric
\eqref{our} which we propose is that it extends through the horizon can therefore be used to study chiral
dynamics near the phase transition. Other proposed metrics written in Eddington-Finkelstein coordinates \cite{Heller08,Kinoshita08} also extend across the horizon, however our metric has the convenient property that it is manifestly AdS as $z\to 0$, which is makes it easier to apply the AdS/CFT prescription.

While we did not obtain
\eqref{our} by solving Einstein's equations in a controlled
expansion, although that is presumably possible\footnote{A possible starting point would be to seek solutions
of Einstein's equations of motion with a metric ansatz
$
ds^2=\frac 1{z^2}\left[-d\tau^2+\left(dz-A(s,\tau)d\tau\right)^2+B(s,\tau)\tau^2 d\eta^2+C(s,\tau)dx_\perp^2\right].$
},
we adopt it as the simplest metric with which to study
non-equilibrium properties of the chiral phase transition in an expanding cooling plasma.  In particular, we will use this background to numerically simulate the evolution of a probe D7-brane `pulling out' of the black hole.

\section{D7-brane Embeddings: the Dynamic Case}\label{dynamicsect}

To describe the dynamical passage of a cooling quark-gluon plasma through the chiral phase transition point, we will embed a D7-brane into the geometry described by the metric \eqref{our}. Simulating the evolution of D7-branes in this background is not an easy task. As will be seen below, the equations of motion are nonlinear with nontrivial dependence on coordinates. Simulations of realistic scenarios will require a serious research in numerical methods and will be attempted elsewhere. Here, we consider a simplified scenario and test to see if the model at hand gives sensible results, both theoretically and numerically. Specifically, we show that a (numerical) solution exists in which the D7-brane pulls out of the horizon, indicating the chiral phase transition. We will discuss our numerical methods and simplifications in detail, hoping that it will benefit future attempts.

Like the AdS black hole case, the D7-brane is assumed to fill the five dimensional geometry and wrap an $S_3$ inside the $S_5$. The brane ends when $S_3$ shrinks to zero size. The three angular coordinates of $S_3$ and the five coordinates of the metric \eqref{our} are chosen to be the parameters that describe the D7-brane embedding.  Throughout this section, we will call the time-like direction $\tau$ the time. Using the previously given form of the five-sphere metric \eqref{five}, we assume  $\phi=0$ and $\theta=\theta(\tau,z)$. This does not take into account anisotropies formed in the dual plasma, in a more realistic simulation one would consider dependence on $\vec x_\perp$ and $\eta$. The asymptotic behavior of the $\theta(\tau,z)$ field as $z\to0$ will be interpreted in the AdS/CFT sense. The quark mass $M$ and chiral condensate $C$ of the dual plasma will again be given by asymptotic behavior of the scalar field, but now they will be time dependent.

We set $e_0=4/3$. The induced metric on the D7-brane is
\begin{align}\label{D7m}
  ds^2 = &\frac 1{z^2} \left[-(1-\tau^{-\frac43}z^4-z^2\dot\theta^2)d\tau^2 +2(z^2\dot\theta\theta'-\tau^{-\frac23}z^2)d\tau dz+(1+z^2\theta'^2)dz^2+\tau^2d\eta^2+d\vec{x}^2_\bot\right]\nonumber\\
  &+ \cos^2\theta dS_3^2.
\end{align}
The D7-brane action, up to some normalization factor is,
\begin{equation}
  S_{D7} = \int dz d\tau \, \frac {\tau}{z^5} \cos^3\theta \sqrt{1+z^2\theta'^2\left(1-\tau^{-\frac 43}z^4\right)-z^2\dot{\theta}^2-2\tau^{-\frac 23}z^4\theta'\dot{\theta}}.
\end{equation}
The equation of motion, after some simplifications, becomes
\begin{align}\label{DynEOM}
&3 \tau^3 z^2 \ddot{\theta}  +3\tau^3 z^4\ddot{\theta} {\theta'}^2 +\left(3\tau^{5/3}z^6-3\tau^3 z^2\right) \theta'' + 3\tau^3 z^4 \dot{\theta}^2\theta''+\left(3 \tau^2 z^2-3 \tau^{7/3} z^3\right) \dot{\theta } \nonumber \\
&+\left(3 \tau^{5/3}z^5+\tau^{4/3}z^4+9 \tau^3z\right)\theta' + \left(6 \tau^{7/3} z^5 -3 \tau^2 z^4 \right)\dot{\theta }^3\nonumber \\
&+\left(6 {\tau }^{1/3} z^{11}-3z^{10}-18 \tau ^{5/3}z^7+\tau^{4/3} z^6+12 \tau^3 z^3\right) {\theta'}^3+\left(18 \tau ^{5/3} z^7-9 \tau^{4/3}z^6-12 \tau^3 z^3\right) \dot{\theta }^2 \theta'  \nonumber \\
&+\left(18 \tau z^9-9 \tau ^{2/3} z^8-36 \tau ^{7/3} z^5+3 \tau ^2 z^4\right)\dot{\theta} {\theta'}^2+ 6\tau^{7/3}z^4\dot{\theta}'-6 \tau^3 z^4 \dot{\theta} \theta' \dot{\theta}' \nonumber \\
&- 9\tau^3\tan{\theta}+9 \tau ^3 z^2\dot{\theta}^2 \tan{\theta} +18 \tau^{7/3} z^4 \dot{\theta}{\theta'}\tan {\theta}+\left(9\tau ^{5/3} z^6-9\tau^3z^2\right){\theta'}^2 \tan{\theta}=0.
\end{align}
This is a second order nonlinear partial differential equation. Polynomials of derivatives  to third order appear, the tangent of $\theta$ contains polynomials of $\theta$ to all orders. The coefficients are both space and time dependent. We assume that the Cauchy problem is well-defined for this equation, in the sense that for some set of initial conditions defined on a space-like hypersurface, the equation will have a unique solution for the causal future. One should not expect this statement to hold for every initial condition, we propose consistency requirements on the initial condition below.

In the considered scenario, an embedding of the D7-brane is specified on an initial time slice. This embedding is chosen to be an infalling one, corresponding to the high temperature, chiral symmetric phase. As the static D7-brane embedding in Lorentzian signature AdS black hole background, considered D7-brane embeddings will end at some point inside the horizon. Furthermore, boundary conditions are set at the boundary of the space at $z=0$, the necessity of this will be apparent below. Then the embedding is iterated in time, during which the plasma is cooling. Geometrically, this corresponds to the horizon moving away from the boundary. We expect to see the D7-brane embedding to be ``pulling out'' of the horizon, through the mechanism discussed above, and at some point becoming non-infalling. This will be interpreted as the dynamical passage through the chiral phase transition.

We start by writing equation \eqref{DynEOM} in first order form. After many trials, we found the following form to be numerically more stable. First we define the variables,
\begin{equation}
  u = \dot\theta, \qquad v = \theta',
\end{equation}
and coefficient functions
\begin{align}
  a &= 3\tau^3z^2+3\tau^3z^4v^2, \nonumber \\
  b &= 6\tau^{7/3}z^4-6\tau^3z^4uv, \nonumber \\
  c &= 3\tau^{5/3}z^6-3\tau^3z^2+3\tau^3z^4u^2, \nonumber \\
  d &= \left(3 \tau^2 z^2-3 \tau^{7/3} z^3\right)u +\left(3 \tau^{5/3}z^5+\tau^{4/3}z^4+9 \tau^3z\right)v\nonumber \\
    &+ \left(6 \tau^{7/3} z^5 -3 \tau^2 z^4 \right)u^3+\left(6 {\tau }^{1/3} z^{11}-3z^{10}-18 \tau ^{5/3}z^7+\tau^{4/3} z^6+12 \tau^3 z^3\right) v^3\nonumber \\
    &+\left(18 \tau z^9-9 \tau ^{2/3} z^8-36 \tau ^{7/3} z^5+3 \tau ^2 z^4\right)u v^2 +\left(18 \tau ^{5/3} z^7-9 \tau^{4/3}z^6-12 \tau^3 z^3\right) u^2 v  \nonumber \\
    &- 9\tau^3\tan{\theta}+9 \tau ^3 z^2u^2 \tan{\theta} +18 \tau^{7/3} z^4uv\tan {\theta}+\left(9\tau ^{5/3} z^6-9\tau^3z^2\right)v^2 \tan{\theta}.
\end{align}
Then, equation of motion \eqref{DynEOM} can be written as
\begin{align}\label{hyper}
\left[ \begin{array}{c} \dot \theta \\ \dot u \\ \dot v \end{array} \right] + \begin{bmatrix} -u/v& 0 & 0 \\ d/(va) &b/a & c/a \\ 0& -1 & 0 \end{bmatrix} \left[\begin{array}{c} \theta' \\ u' \\ v' \end{array}\right] = \bold 0.
\end{align}
Written in this form the equation is quasi-linear with no source term. The first component equation merely states a choice of integrating $\theta$, it actually is a constraint equation. Through trial and error we found the above form to be numerically more stable. The second and the third component equations contain the real information of the equation of motion, they relate the second derivatives of $\theta$. To classify this system as an hyperbolic, an elliptic or a parabolic system of equations, one has to solve for the eigenvalues  and eigenvectors of the 3-by-3 matrix appearing on the left of equation \ref{hyper}, \cite{DuChateau02}. This matrix depends on the solution itself, as well as independent variables, therefore it is not possible to do a classification before obtaining a solution, which is crucial in posing the correct problem and choosing the numerical method. On the other hand, the physics of the problem asks for an hyperbolic equation, in which the information of the initial time embedding determines the embedding in later times (boundary conditions will also be needed as will be discussed below). Therefore, we assume that the equation is hyperbolic. This choice puts strict consistency conditions to the numerical solution. At every point in space and time, the 3-by-3 matrix appearing on the left of equation \eqref{hyper} should have real eigenvalues (characteristic velocities) with linearly independent eigenvectors. The physics of the problem tightens this restriction. Within the horizon, we expect the information flow to be only in positive $z$ direction pointing towards the singularity. Going back to the equation \eqref{hyper}, we see that one of the eigenvalues of the 3-by-3 matrix appearing on the left will always be given by $-u/v$, corresponding to the constraint equation. The remaining two eigenvalues ($b/2a \pm \sqrt{b^2-4ac}/2a$) come form the second and the third equations, which relate the second derivatives of $\theta$. These eigenvalues then must be real and inside the horizon they must be positive. In terms of the variables defined above, the reality condition for eigenvalues is (for real $\theta$, $u$ and $v$)
\begin{align}\label{reality}
b^2-4ac>0 \implies -z^6v^2-2\tau^{\frac 23}z^4uv+\tau^{\frac 43}(1-u^2z^2+v^2z^2)>0.
\end{align}
For the positivity constraint, the additional condition is
\begin{align}\label{positivity}
ac>0 \implies z^4+\tau^{\frac 43}(-1+z^2u^2)>0.
\end{align}
In simplifying these equations, we made use of the positivity of $\tau$ and $z$. Recalling that the horizon is located at $z_h(\tau)=\tau^{1/3}$, the equation \eqref{positivity} is automatically satisfied inside the horizon for any embedding. This confirms our expectations. These restrictions should be taken into account when choosing an initial condition.

We use a simple first order upwind scheme to integrate the set of equations \eqref{hyper}. Upwind scheme is introduced in the appendix \ref{Upwind}.

Having specified the scheme, we now discuss the initial and boundary conditions required for solving the system of equations \eqref{hyper}. Aside from specifying a D7-brane embedding at some initial time, the scheme \eqref{upwindscheme} requires boundary conditions at the boundaries of the domain to propagate information. This is not just a numerical necessity, the boundary conditions carry information from the causal past of the solution that is not included in the initial condition. Recall that we consider the scenario in which the initial condition is an infalling D7-brane embedding. Initially the D7-brane extends from the boundary of the space given by the hypersurface $z=0$ to its endpoint $z_{\text{end}}$ which is inside the horizon. One needs to set boundary conditions at the spatial boundary, $z=0$,  to accommodate for incoming information. The functions $\theta$, $u$ and $v$ will be set at $z=0$ and this will be fed to the backward differenced term in the upwind scheme \eqref{upwindscheme} at the first grid point to be propagated in the positive direction. For our numerical simulation, the boundary condition at $z=0$ is chosen to be
\begin{align}
	\theta(\tau,0) = 0, \qquad u(\tau,0) = 0, \qquad v(\tau,0) = M\tau_i^{1/3},
\end{align}
where $M$ is a constant quark mass up to some normalization. Note that the temperature goes like $T\sim \tau^{-1/3}$, hence the $\tau^{1/3}$ factor is included in the statement above.
%For realistic scenarios, one may need to consider a time dependent quark mass \cite{Randrup01}.

At the other hand of the brane, we do not expect a need for a boundary condition. In the expected evolution, the endpoint of the brane moves towards the boundary of space, ``pulling out" of the horizon. The location of the endpoint changes over time, therefore the domain of the solution that we are looking for is dynamically being updated at every time step. We expect the endpoint of the brane to be determined only by the brane configuration in the previous time step. For infalling embeddings, due to our expectancy of characteristic speeds being positive inside the horizon there is no numerical necessity for a boundary condition at the endpoint. For a non-infalling embedding, we chose to ignore negative characteristic speeds at the endpoint and only propagate positive characteristic speeds. We suggest an alternative approach below, see footnote $^{\ref{footnote}}$. Even though one does not expect the necessity of a boundary condition at the endpoint, there is a natural consistency condition that comes from the equation of motion \eqref{DynEOM}. The brane ends when $S_3$ shrinks to zero size. An inspection of the induced metric on the brane \eqref{D7m} shows that this happens when $\theta = \pi/2$. The point at which the brane ends evolves in time, $z_{\text{end}}=z_{\text{end}}(\tau)$. However, $\theta(\tau,z_{end}(\tau))=\pi/2$. Then
\begin{align}\label{totalder}
  \left.\frac{d\theta}{d\tau}\right|_{z=z_{\text{end}}(\tau)}=\theta'_{\text{end}}\dot z_{\text{end}}+\dot\theta_{\text{end}}=0,
\end{align}
where
\begin{align}
\theta_{\text{end}}'=\left.\frac{\partial \theta}{\partial z}\right|_{z=z_{\text{end}}(\tau)}, \qquad \dot\theta_{\text{end}}=\left.\frac{\partial \theta}{\partial \tau}\right|_{z=z_{\text{end}}(\tau)}.
\end{align}
For an infalling embedding, we expect $\theta$ and its first and second derivatives to be nonsingular at the endpoint. The most  singular terms of the equation of motion \eqref{DynEOM} are those that are proportional to $\tan \theta$. Requiring the coefficients of the most singular terms to cancel leads to
\begin{align}\label{endpointequation}
 1= z_{end}^2\dot{\theta}_{end}^2+\tau^{-4/3} z_{end}^6 {\theta'}_{end}^2 + 2 \tau^{-2/3} z_{end}^4 \dot{\theta}_{end}{\theta'}_{end}-z_{end}^2{\theta'}_{end}^2.
\end{align}
Combining these equations, one can solve for $z_{\text{end}}(\tau)$,
\begin{align}\label{zenddot}
  \dot{z}_{end} = -\tau^{-2/3}z_{end}^2 \pm \sqrt{1+\frac{1}{{\theta'}_{end}^2z_{end}^2}}.
\end{align}
We use equations \eqref{endpointequation} and \eqref{zenddot} to check the consistency of our numerical solution. For a non-infalling embedding $\theta$ is expected to be singular, so that the brane closes smoothly. To get a constraint on the endpoint one needs to know the exact asymptotic behavior of the embedding\footnote{\label{footnote}Being inspired by the AdS black hole case, we propose the following form of the solution, for $z$ near $z_{\rm end}$ for an non-infalling embedding
\begin{align}
\theta(\tau,z) \sim \frac{\pi}{2} -  \left(z_{\rm
end}(\tau)-z\right)^{1/2}\sum_{n=0} \alpha_n(\tau)(z_{\rm end}(\tau)
-z)^n
\end{align}
Although we have not used this proposal in our numerical simulations, in further studies this form could be used to propagate the endpoint by solving the equation of motion order by order for $z$ near $z_{\text{end}}$.}.

\begin{figure}[ht]
\begin{center}
\includegraphics[width=\textwidth]{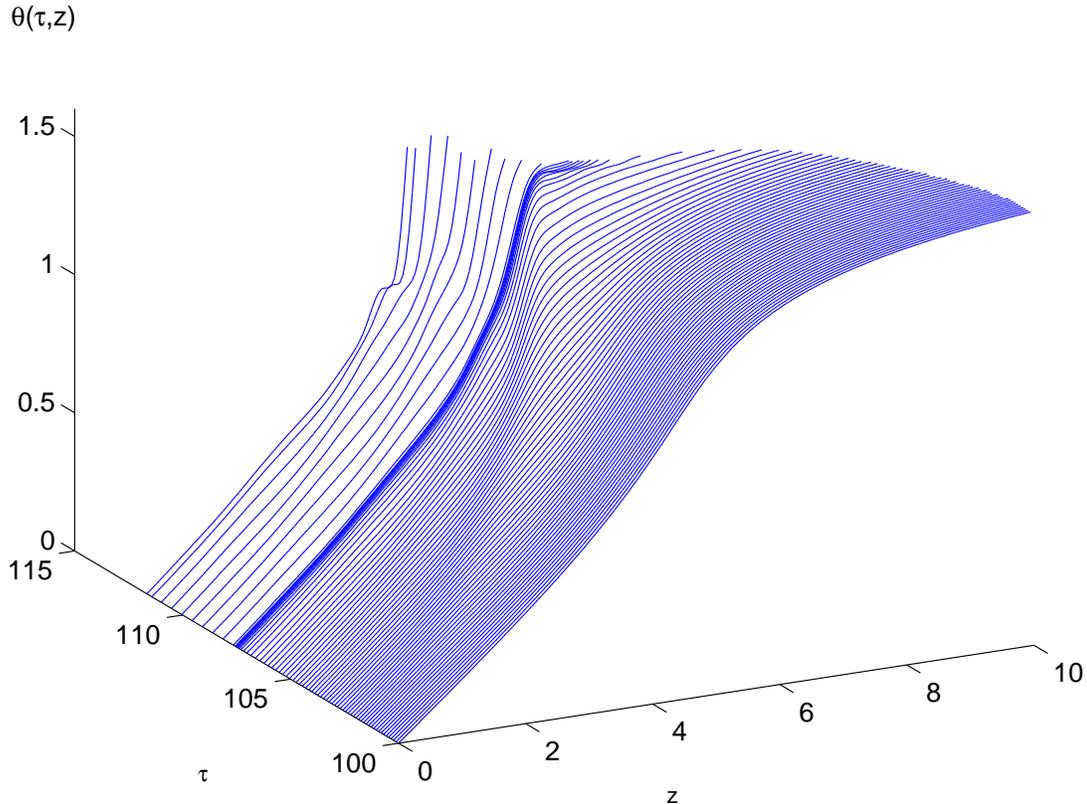}
\end{center}
\caption{\label{simulation1} The evolution of a D7-brane with the inital condition discussed in the text.}
\end{figure}

\begin{figure}[ht]
\begin{center}
\includegraphics[width=\textwidth]{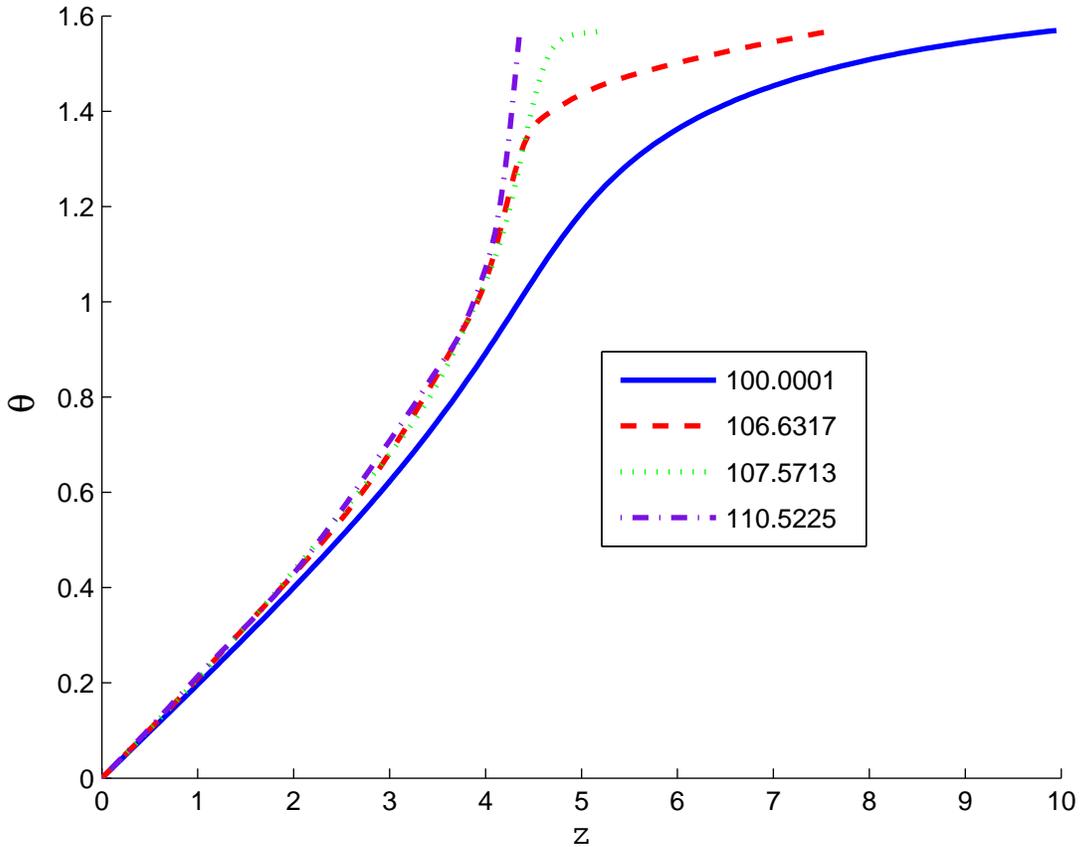}
\end{center}
\caption{\label{simulation2} The embedding at different time slices. }
\end{figure}

\begin{figure}[ht]
\begin{center}
\includegraphics[width=\textwidth]{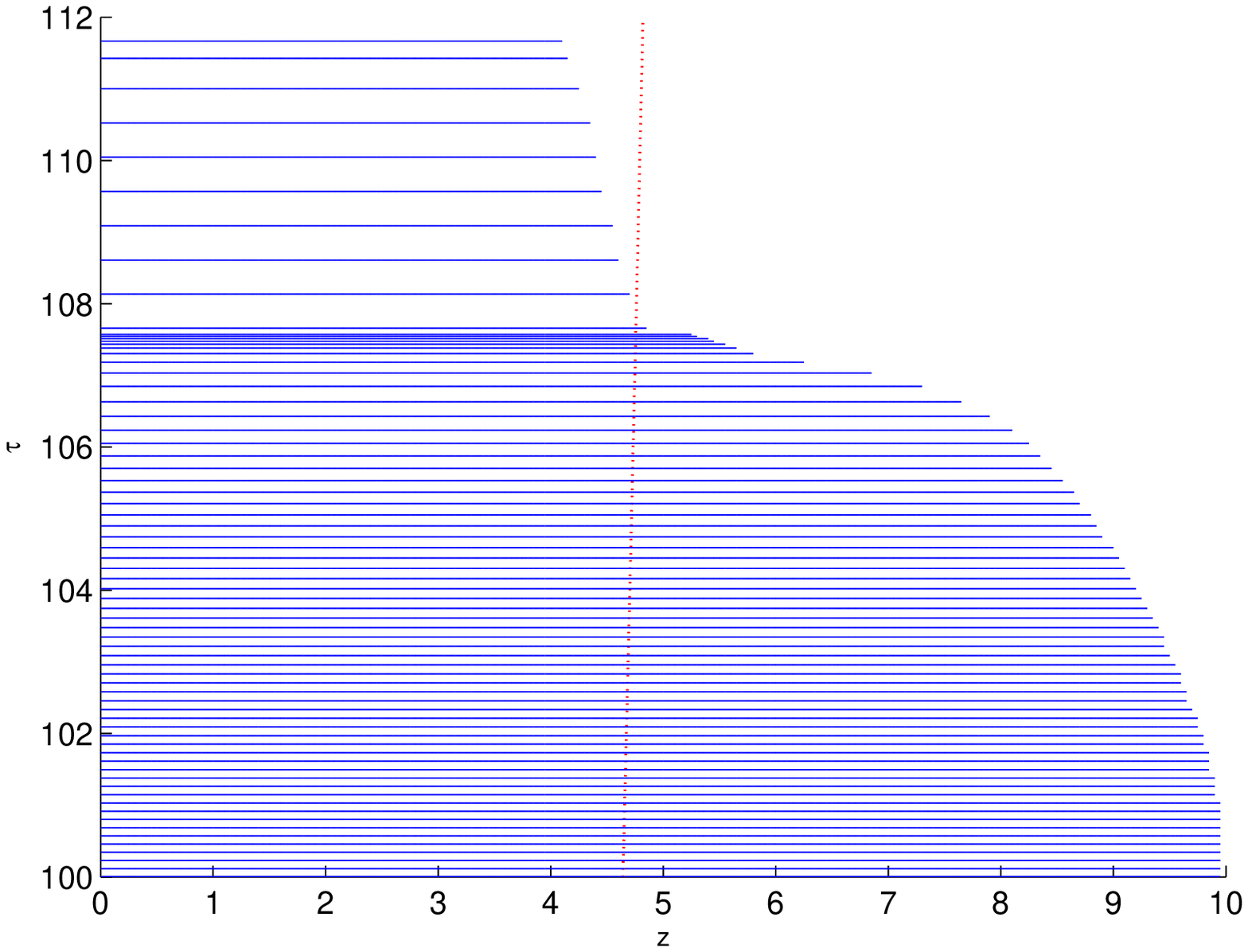}
\end{center}
\caption{\label{simulation3} The same evolution with $\theta$ direction pointing outside the paper plane. In this figure, the red (dashed) line shows the evolution of the black hole horizon.}
\end{figure}

We choose an infalling embedding as an initial condition. This choice is subject to consistency conditions. First of all, it should be compatible with the boundary conditions. Moreover, it should obey the restrictions on characteristic velocities discussed above. The eigenvalues of the 3-by-3 matrix appearing on the left of equation \eqref{hyper} should have real eigenvalues with linearly independent eigenvectors at every point on the grid, which requires the condition \eqref{reality}. The condition of two of the three eigenvalues corresponding to second and third equations of the system \eqref{hyper} being positive inside the horizon is automatically satisfied, as discussed below equation \eqref{positivity}. The asymptotics of the embedding near the endpoint should satisfy \eqref{endpointequation}. It is not trivial to find an initial condition that satisfies all these properties. Our starting point was noticing that when $u=0$, the constraint equation \eqref{endpointequation} on the brane endpoint is satisfied by an infalling D7-brane embedding in an AdS black hole background with static horizon at $z_h=\tau_i^{1/3}$. We chose such an embedding, which sets $\theta$ and $v$ at the initial time slice, and $u=0$ as initial conditions. For consistency, one should look for a D7-brane embedding with quark mass that is equal to the boundary condition for the whole evolution (which we chose to be static). This is numerically done by a shooting method. Moreover, we numerically verified that the reality of characteristic velocities condition \eqref{reality} is also satisfied for this initial condition choice. Figures \ref{simulation1}, \ref{simulation2} and \ref{simulation3} show the results for a sample simulation. Some additional special techniques were used to stabilize the system, which are described below. We were able to numerically integrate the D7-brane evolution into a non-infalling configuration. After some point numerical instabilities were not controllable, whose initial stages can be seen in the figure.\footnote{Another set of initial conditions could be given by setting $\dot\theta$, ${\ddot\theta}$ and $\dot\theta'$ to zero at some initial time $\tau_i$, which we call the ``static initial condition'' (although the embedding appears static only momentarily and evolves in time), and solve for the embedding at this constant time slice. The equation of motion \eqref{DynEOM} gives
\begin{align}\label{DynEOMStat}
	\theta''& = \frac{3\tan\theta}{z^2\left(\tau^{-4/3}_iz^4-1\right)} - \frac{\tau^{-4/3}_iz^4+3}{z\left(\tau^{-4/3}_iz^4-1\right)}\theta' - \frac{\tau^{-5/3}_iz^2}{3\left(\tau^{-4/3}_iz^4-1\right)}\theta'- 3\tan\theta\theta'^2\nonumber \\
	&\qquad  + 2z(2-\tau^{-4/3}_iz^4)\theta'^3 + \frac{3z^8-\tau^{4/3}_iz^4}{3\tau^{5/3}_iz^4-3\tau^3_i}\theta'^3.
\end{align}
This is to be contrasted with AdS black hole embedding \eqref{staticEOM}. One sees that third and the sixth terms on the right hand side are new. Let's note that in the $z\to z_{\text{end}}$ limit, the static initial condition does satisfy the consistency condition \eqref{endpointequation}. This actually is a redundant statement since the static embedding is solved from the equation of motion anyway, as equation \eqref{endpointequation} also is. To find an infalling solution for this equation, we follow the same line of argument as in the AdS black hole case. The horizon is at $z_h(\tau)=\tau^{1/3}$. We choose a value for $\theta$ at the horizon that is less than $\pi/2$, $\theta_0<\pi/2$. Then, from the equation above, one can solve for the condition on $\theta'_0$ requiring the coeffiecients of the leading order singularity to cancel,
\begin{align}\label{root}
	\frac 23 \tau^{-1/3}_i\theta_0'^3-\frac 13 \tau^{-1}_i\theta_0'-4\tau^{-1/3}_i\theta_0'+3\tau^{-2/3}_i\tan\theta_0 = 0.
\end{align}
For an infalling embedding, we expect to have a positive slope at the horizon. The reality and positivity of roots depend on particular $\tau_i$. As in the AdS black hole case, the equation \eqref{DynEOMStat} is numerically integrated both towards the singularity and the boundary starting from the horizon with boundary conditions $\theta_0$ and $\theta'_0$. We found that not all positive real roots of equation \eqref{root} led to boundary conditions that were numerically integrable. We tried this setting for a various $\theta_0$ and $\tau_i$ and verified numerically the existence of initial condition that satisfies the reality condition of characteristic velocities \eqref{reality}. The time evolution of the corresponding initial conditions also led to topology changing D7-brane evolutions, satisfying the condition on the endpoint evolution \eqref{endpointequation} within numerical accuracy.}

In the figures, the evolution starts from an infalling configuration at $\tau_i = 100$ and proceeds to a non-infalling configuration. The horizon crosses the endpoint at $\tau=107.8003$. After some point numerical instabilities lead to uncontrollable oscillations and the numerical integration fails. The initial stages of the formation of these oscillations can be seen in the top figure. The quark mass is set to $M\tau_i^{1/3}=0.1942$. Grid spacing in the $z$ coordinate is $\Delta z = 0.05$. Numerical integration is done until the last point of the D7-brane that is on a grid point. Adaptive time steps are chosen to be $0.01$ times the maximum time step allowed by the CFL condition \eqref{CFL}. In the top and bottom figures we only present 1 time step in every 1000 steps. The embedding lines become denser where the numerical integration proceeds by smaller step sizes. This was not sufficient to prevent the instability that discretization induced on the system. Unstable oscillations grew quickly in time and failed the numerical integration. To prevent this we introduced a sixth order sponge filter, see \cite{Alcubierre08} for a discussion of sponge filters. Near $z=0$, we used a polynomial fit to smooth out the unstable oscillations. To check the accuracy of the numerical solution, we checked the condition on the endpoint \eqref{endpointequation} at every time step at the last grid point for the infalling configuration. The absolute value of the difference between the left side and the right side was always much below the spatial grid size. The error increased when the grid point got away from the actual endpoint. For the infalling configuration, the absolute value of the error when averaged over all time steps was $0.0002$, which we interpreted as a confirmation of the numerical result.


\section{Conclusions}

We have made some initial steps toward describing the non-equilibrium dynamics of a first order phase transition akin the chiral phase transition in QCD, in a strongly coupled supersymmetric large $N_c$ gauge theory,  by making use of AdS/CFT duality. The dual description of the phase transition involves a probe D-brane falling into or 'pulling out' of an AdS-black hole.

One aspect of this work involved an attempt to obtain a Lorentzian signature description of the latent heat.  The probe D-brane geometry in the infalling case implies a lower bound on the energy flux at the boundary of Anti-de-Sitter space (and at the horizon) in crossing from the high temperature phase to the low temperature phase.  However this bound was not stringent enough, being less than the actual latent heat computed in Euclidean signature.  It would be interesting to resolve this discrepancy, which presumably involves an accounting for the back-reaction.

We have also described our initial attempts at simulating the phase transition in a background dual to a boost invariant expanding cooling plasma, akin to that produced in heavy ion collisions.  Our proposed background is of the Painlev\'e-Gullstrand, or 'river-model' form which seems to be particularly convenient, giving a description valid across the horizon and which is manifestly AdS asymptotically.  The numerical simulation of the highly non-linear equations of motion in a space-time with a complicated causal structure is non-trivial.  We have restricted ourselves to the simplest case of supercooling, without introducing spatial inhomogeneities.  In our simulations, we observed a D7-brane 'pulling out' of the black hole, however we encountered instabilities shortly after the transition.  Further refinements are clearly needed to mollify the numerical instability.

\section*{Acknowledgments}

G. Guralnik and C. Pehlevan were supported in part by Grant JSF 09020001 from the Julian Schwinger Foundation.
They were also supported in part by the US Department of Energy under DE-FG02-91ER40688-TaskD.
We would like to thank Richard Brower and Fred Cooper for participating in earlier unpublished work on theoretical aspects related to this project.

\appendix

\section{Upwind Scheme}\label{Upwind}
Upwind scheme is a method for solving hyperbolic partial differential equations. In this scheme, the partial differential equation is discretized by using differencing biased in the direction determined by the sign of the characteristic speeds. The direction of propagation of information is taken into account through this mechanism. We refer the reader to literature for theoretical results on the upwind scheme. We found the discussions in \cite{DuChateau02,Gustafsson96,Mattheij05,Morton05,Langtangen99,Alcubierre08} to be very useful. Here we briefly describe the method, using the notation of \cite{Mattheij05}.

Given a system of first order partial differential equations in two independent variables,
\begin{align}
  \frac {\partial \bold y}{\partial t } + \bold B(t,x,\bold y) \frac{\partial \bold y}{\partial x} = \bold f(t,x,\bold y),
\end{align}
one first introduces a grid over the domain $[0,\infty)\times[0,\infty)$ with grid points $(t^n,x_j)$ defined by
\begin{align}
  x_j := j\Delta x \qquad (j=0,1,2,\ldots), \qquad t^n = \sum_{k=0}^{n-1}\Delta t^k \qquad (n=0,1,2,\ldots).
\end{align}
with $\Delta x$ the spatial grid size and $\Delta t^k$ the time step. The time step is adaptive, it will be defined at each step by the Courant-Friedrichs-Lewy (CFL) condition as discussed below. The matrix $\bold B$ is assumed to have real eigenvalues and a complete set of eigenvectors at every point over the domain for the particular $\bold y$ in question; this is the condition of hyperbolicity. The upwind scheme is given by
\begin{align}\label{upwindscheme}
\frac{1}{\Delta t^n}(\bold y_j^{n+1}-\bold y_j^n)+\frac{1}{\Delta x}\bold B_j^{n,+}(\bold y_j^n-\bold y_{j-1}^n) + \frac 1{\Delta x}\bold B_j^{n,-}(\bold y^n_{j+1}-\bold y_j^n) = \bold f(x_j,t^n,\bold y_j^n).
\end{align}
The scheme is solved iteratively for $\bold y_j^{n+1}$. The matrices $\bold B_j^{n,+}$ and $\bold B_j^{n,-}$ are defined locally by
\begin{align}
\bold B_j^{n,+} := \bold S_j^{n}  \bold \Lambda_j^{n,+} (\bold S_j^{n})^{-1}, \qquad \bold B_j^{n,-} := \bold S_j^{n}  \bold\Lambda_j^{n,-}(\bold S_j^{n})^{-1}.
\end{align}
$\bold S_j^{n} $ is the similarity transformation matrix that consists of the local eigenvectors of $\bold B_j^{n} $ as its columns, i.e. if $\bold b_{j,1}^{n} $, $\bold b_{j,2}^n$, and $\bold b_{j,3}^n$ are a complete set of (right) eigenvectors of $\bold B_j^n$ for the grid point $(t^n,x_j)$ and the value of the function $\bold y$ at the grid point, $\bold y_j^n$, then $\bold S_j^n = (\bold b_{j,1}^n, \bold b_{j,2}^n, \bold b_{j,3}^n)$. $\bold \Lambda_j^{n,+}$ and $\bold \Lambda_j^{n,-}$ are diagonal matrices that contain the positive and negative eigenvalues of the matrix $\bold B_j^n$ as diagonal elements respectively, i.e. if $\lambda_{j,1}^n$, $\lambda_{j,2}^n$ and $\lambda_{j,3}^n$ are the eigenvectors corresponding to $\bold b_{j,1}^n$, $\bold b_{j,2}^n$, and $\bold b_{j,3}^n$, then $(\bold \Lambda_j^{n,+})_{ii}= \max(\lambda_{j,i}^n,0)$ and $(\bold\Lambda_j^{n,-})_{ii}= \min(\lambda_{j,i}^n,0)$, $i=1,2,3$. $\bold B_j^{n,+}$ and $\bold B_j^{n,-}$ give a decomposition of $\bold B_j^{n}$, $\bold B_j^{n} = \bold B_j^{n,+} +\bold B_j^{n,-}$. We note again that this decomposition has to be done at every grid point and will be different for different functions $\bold y$.

The eigenvalues $\lambda^n_{j,1}$, $\lambda^n_{j,2}$ and $\lambda^n_{j,3}$ are the local characteristic velocities of the system, they describe the direction and speed of information flow. The upwind scheme \eqref{upwindscheme} uses a backward differencing to propagate information in the positive direction and a forward differencing to propagate information in the negative direction. The time stepping should be adaptively chosen by the CFL condition,
\begin{align}\label{CFL}
\max_{j,i}(\lambda_{j,i}^n)\frac{\Delta t^n}{\Delta x}<1,
\end{align}
for stability \cite{DuChateau02,Gustafsson96,Mattheij05,Morton05,Langtangen99,Alcubierre08}. This condition prevents acausal information flow on the grid.

The upwind scheme is dissipative. There are many other schemes with
better accuracy, but their implementation to the equation at hand is
very difficult due to the nonlinearity and parameter dependent
coefficients.

\bibliographystyle{plain}
\bibliography{paper}

\end{document}